\documentclass[preprint,prd,showpacs]{revtex4} 
 
\begin{document} 
\input{epsf}


\title{The Effects of Stress Tensor Fluctuations upon Focusing}

\author{J. Borgman}
\author{ L.H. Ford} 
 \email[Email: ]{ford@cosmos.phy.tufts.edu} 
 \affiliation{Institute of Cosmology  \\
Department of Physics and Astronomy\\ 
         Tufts University, Medford, MA 02155}

\begin{abstract} 
 We treat the gravitational effects of quantum stress tensor fluctuations.
An operational approach is adopted in which these fluctuations produce
 fluctuations in the focusing of a bundle of geodesics. This can be calculated
explicitly using the Raychaudhuri equation as a Langevin equation. The
physical manifestation of these fluctuations are angular blurring and
luminosity fluctuations of
the images of distant sources. We give explicit results for the case
of a scalar field on a flat background in a thermal state. 
\end{abstract}

\maketitle 
 
\baselineskip=13pt 

\section{Introduction}

Although quantum states in field theory may or may not be eigenstates 
of the Hamiltonian,
and hence have fixed energy, they are never eigenstates of the energy
density or other stress tensor components. This means that all quantum
field theories exhibit fluctuations of the stress tensor. These fluctuations
can manifest themselves either through fluctuating forces on material bodies,
or through fluctuations of the gravitational field. In the former category
are fluctuations of Casimir forces~\cite{Barton,Eberlein,JR,WKF01} and
fluctuations of the radiation pressure~\cite{WF01}. In this paper, we will
be concerned with the latter phenomenon, where stress tensor fluctuations
drive fluctuations of the spacetime geometry. These are ``passive''
fluctuations of the spacetime, in contrast to the ``active'' fluctuations
which arise from the quantization of gravity itself. The role of 
stress tensor fluctuations in gravity theory has been discussed by
several authors~\cite{F82,DF88,Kuo,PH97,WF99,CH95,CCV,MV99,HS98,BFP00,PH00,FW03}.
Among the issues of interest is the extent to which these fluctuations
place limits on the validity of the semiclassical theory of gravity, which is 
the approximation in which the gravitational field is assumed to be classical
and determined by the renormalized expectation value of the stress 
tensor~\cite{F82,Kuo,PH00}. Other authors have been concerned with the possible
role of stress tensor fluctuations in the early universe or upon black
hole evaporation~\cite{CH95,CCV,MV99,HS98,WF99,BFP00}.

One of the basic problems which must be addressed in treatments of quantum
stress tensor fluctuations is that the stress tensor correlation function is
singular, even if the stress tensor operator has been renormalized to have
finite expectation values. For this reason, many of the earlier 
treatments~\cite{F82,DF88,Kuo,PH97,WF99} studied only a portion of the 
correlation function which is free of singularities, the ``fully normal
ordered part''. However, more recently it has become clear that the remaining
parts (the ``vacuum'' and ``cross'' terms to be defined in 
Sect.~\ref{sec:QSTF}) can also have physical effects,
and that extracting these effects requires a careful procedure for defining
integrals with singular integrands~\cite{WF01,FW03}. For example, the
radiation pressure fluctuations, arising when light in a coherent state impinges
upon a mirror, come entirely from the cross term~\cite{WF01}. 

In this paper, we will present an approach to studying the gravitational effects
of stress tensor fluctuations which is based upon Raychaudhuri's equation.
This equation relates the Ricci tensor to the rate of change of the
expansion along a congruence of geodesics, and plays a central role in
the proofs of the classical singularity theorems. We will treat it as
a Langevin equation which describes how stress tensor fluctuations lead
to fluctuations in the expansion. This in turn leads to a lack of focusing and fluctuations in the luminosity of the images of distant objects seen through a fluctuating
spacetime geometry. Here the quantum matter field will be a massless,
minimally coupled scalar field on an average background of Minkowski
spacetime. The outline of the paper is as follows: 
Section~\ref{sec:QSTF} will discuss the stress tensor correlation function and
its decomposition into individual terms. Section~\ref{sec:ricci} will
deal with the Ricci tensor correlation function and the fluctuations of the
expansion. The treatment of the singular pieces in the correlation function will
also be discussed in this section. In Sec.~\ref{sec:thermal}, we will 
explicitly calculate the expansion fluctuations for a thermal state. The results
of the paper will be summarized and discussed in Sec.~\ref{sec:final}.
Except as otherwise noted, we use units in which $\hbar = c = G = k =1$,
where $G$ and $k$ are Newton's and Boltzmann's constants, respectively.

\section{Quantum Stress Tensor Fluctuations}
\label{sec:QSTF}

In this paper, we will deal with the fluctuations of the stress tensor for
quantum fields on an average background of
Minkowski spacetime. 
The formal expectation values of $T_{\mu\nu}$ are divergent, so we
renormalize by subtracting the expectation value in the Minkowski vacuum
state:
\begin{equation} 
:\!T_{\mu \nu }\!:=T_{\mu \nu }-\langle T_{\mu \nu }\rangle_{0}\, ,
\end{equation}
where $:\!T_{\mu \nu }\!:$ is the normal ordered stress tensor operator.
The fluctuations of $:\!T_{\mu \nu }\!:$ are described by the stress tensor 
correlation function
\begin{equation}
S_{\mu \nu \alpha \beta}(x,x') = 
\langle :\!T_{\mu \nu }(x)\!: \, :\!T_{\alpha \beta}(x')\!: \rangle
- \langle :\!T_{\mu \nu }(x)\!: \rangle 
\langle :\!T_{\alpha \beta}(x')\!: \rangle \,.
\end{equation}
Here the expectation value can be taken in an arbitrary quantum state.
It is of interest to note that this correlation function is unchanged when
$:\!T_{\mu \nu }\!:$ is shifted by a constant, so we could formally write
$S_{\mu \nu \alpha \beta}(x,x')$ without normal ordering:
\begin{equation}
S_{\mu \nu \alpha \beta}(x,x') = 
\langle T_{\mu \nu }(x) T_{\alpha \beta}(x') \rangle
- \langle T_{\mu \nu }(x) \rangle \langle T_{\alpha \beta}(x') \rangle \,.
\end{equation}

Although $\langle :\!T_{\mu \nu }(x)\!:\rangle$ is finite in all well behaved
quantum states, $S_{\mu \nu \alpha \beta}(x,x')$ is singular in the limit that
$x$ and $x'$ are null separated. There is a useful decomposition of operator
products which classifies the various contributions to the correlation function.
Let $T(x) = :\!\phi_1(x) \phi_2(x)\!:$, where $\phi_1$ and $\phi_2$ are free 
quantum fields, or derivatives of fields. The stress tensor operator
$:\!T_{\mu \nu }\!:$ is a sum of such terms. It may be shown using Wick's 
theorem that 
\begin{equation}
T(x)\, T(x') = S_0 +S_1 + S_2 \,,  \label{eq:parts}
\end{equation}
where
\begin{equation}
S_0 = 
\langle \phi_1(x) \phi_1(x') \rangle_0 \langle \phi_2(x) \phi_2(x') \rangle_0 
+ \langle \phi_1(x) \phi_2(x') \rangle_0 \langle \phi_2(x) \phi_1(x') \rangle_0
\,,  \label{eq:S0}
\end{equation}

\begin{eqnarray}
S_1 &=& :\!\phi_1(x) \phi_1(x')\!:  \langle \phi_2(x) \phi_2(x') \rangle_0 +
:\!\phi_1(x) \phi_2(x')\!:  \langle \phi_2(x) \phi_1(x') \rangle_0  \nonumber \\
&+& :\!\phi_2(x) \phi_1(x')\!:  \langle \phi_1(x) \phi_2(x') \rangle_0 +
:\!\phi_2(x) \phi_2(x')\!:  \langle \phi_1(x) \phi_1(x') \rangle_0 \,, 
                                 \label{eq:S1} 
\end{eqnarray}
and
\begin{equation}
S_2 = :\!\phi_1(x) \phi_2(x) \phi_1(x') \phi_2(x')\!: \,.  \label{eq:S2}
\end{equation}
Thus the operator product $T(x)\, T(x')$ consists of a purely vacuum part
$S_0$, a fully normal-ordered part $S_2$, and a part $S_1$ which is a
cross term between the vacuum and normal-ordered parts. 

The expectation value of the product of stress tensors can be expressed
as the sum of these three contributions:
\begin{eqnarray}
\langle :\!T_{\mu \nu }(x) \, T_{\alpha \beta}(x')\!: \rangle &=&
\langle :\!T_{\mu \nu }(x)\!: \, :\!T_{\alpha \beta}(x')\!: \rangle +
\langle T_{\mu \nu }(x) T_{\alpha \beta}(x') \rangle_{cross} \nonumber \\
 &+& \langle T_{\mu \nu }(x) T_{\alpha \beta}(x') \rangle_{vac} \, ,
\end{eqnarray}
where the last two terms are the cross term and vacuum term, respectively.
In the limit that $x' \rightarrow x$, 
\begin{equation}
\langle T_{\mu \nu }(x) T_{\alpha \beta}(x') \rangle_{cross} \sim 
\frac{1}{(x-x')^4} \, ,
\end{equation}
and
\begin{equation}
\langle T_{\mu \nu }(x) T_{\alpha \beta}(x') \rangle_{vac} \sim 
\frac{1}{(x-x')^8} \, .
\end{equation}
Observable quantities will be expressed as integrals of the correlation 
function. However, the singularities in the cross and vacuum terms are not
integrable in a naive way. As discussed in Refs.~\cite{WKF01,WF01,FW03}, 
one solution to this problem
is an integration by parts procedure. This method allows one to define the
relevant integrals, and will be discussed in more detail below. An alternative
approach, which will not be pursued in detail in this paper, is to use 
dimensional regularization. In this approach, the divergent part of 
a stress tensor correlation function can be expressed in a local form,
which arises from counterterms in the action of the form of $R^2$
and $R_{\mu\nu} R^{\mu\nu}$, where  $R_{\mu\nu}$ is the Ricci tensor~\cite{RW}.

\section{Ricci Tensor Fluctuations and Raychaudhuri's Equation}
\label{sec:ricci}

\subsection{Raychaudhuri's Equation as a Langevin Equation}

For a classical spacetime, Einstein's equations relate the Ricci tensor
$R_{\mu\nu}$ to the stress tensor by
\begin{equation}
R_{\mu\nu} = 8 \pi \left(T_{\mu\nu} - 
     \frac{1}{2} g_{\mu\nu}\, T^\rho_\rho \right) \,.  \label{eq:Einstein}
\end{equation}
In the case of a spacetime with fluctuations driven by stress tensor
fluctuations, we assume that this algebraic relation between the Ricci
and stress tensors still holds. A useful probe of the fluctuations of the
Ricci tensor is the  Raychaudhuri equation.

Consider a congruence of either timelike or null geodesics with affine
parameter $\lambda$ and tangent vector field $k^\mu$. The  Raychaudhuri 
equation gives the rate of change of  the expansion $\theta$ along the 
congruence to be~\cite{Wald}
\begin{equation}
\frac{d \theta}{d \lambda} = - R_{\mu\nu} k^\mu k^\nu - a\, \theta^2
-\sigma_{\mu\nu} \sigma^{\mu\nu} + \omega_{\mu\nu}  \omega^{\mu\nu} \,.
                                                 \label{eq:ray}
\end{equation}
Here $\sigma^{\mu\nu}$ is the shear and $\omega^{\mu\nu}$ is the vorticity
of the congruence. The constant $a=1/2$ for null geodesics, and $a=1/3$ for 
timelike geodesics. We are interested in interpreting this equation as a 
Langevin equation in which the Ricci tensor fluctuates. For the purposes of 
this paper, we will further assume that the shear and vorticity of the 
congruence vanishes, and that the expansion remains sufficiently small that
the $\theta^2$ term can also be ignored, so we can write
\begin{equation}
\frac{d \theta}{d \lambda}  = - R_{\mu\nu} k^\mu k^\nu \, ,
\end{equation}
where $ R_{\mu\nu}$ is determined by the fluctuating stress tensor by
Eq.~(\ref{eq:Einstein}). 

Let the Ricci tensor correlation function be defined to be
\begin{equation}
C_{\mu \nu \alpha \beta}(x,x') = 
\langle R_{\mu \nu }(x) R_{\alpha \beta}(x') \rangle
- \langle R_{\mu \nu }(x) \rangle \langle R_{\alpha \beta}(x') \rangle \,.
\end{equation}
The variance of the expansion can be expressed as a double integral of this
correlation function as
\begin{equation}
\langle \theta^2 \rangle- \langle \theta \rangle^2 =
\langle (\Delta \theta)^2 \rangle=
\int_0^{\lambda_0} d\lambda \int_0^{\lambda_0} d\lambda' \,
 C_{\mu \nu \alpha \beta}(\lambda,\lambda')\, k^\mu (\lambda) k^\nu (\lambda)
\, k^\alpha(\lambda') k^\beta(\lambda') \,. \label{eq:var}
\end{equation}
Here we are assuming that the fluctuations of the expansion occur
in a finite affine parameter interval, between $\lambda =0$ and 
$\lambda = \lambda_0$. In the case of a flat background, we can take the
stress tensor to be normal ordered. Then the Ricci tensor correlation
function $C_{\mu \nu \alpha \beta}(x,x')$ is related to the correlation
function for the normal ordered stress tensor by
\begin{equation}
C_{\mu \nu \alpha \beta}(x,x') = 64 \pi^2 \left(S_{\mu \nu \alpha \beta}
- \frac{1}{2} g_{\mu\nu}\, {S^\rho}_{\rho \alpha \beta}
- \frac{1}{2} g_{\alpha \beta}\, {S_{\mu \nu \rho}}^\rho
+ \frac{1}{4} g_{\mu\nu}\, g_{\alpha \beta}\, 
                                 {{{S^\rho}_\rho}^\sigma}_\sigma \right) \,,
\end{equation}
and can also be decomposed into fully normal ordered, cross, and vacuum terms.

In the remainder of this paper, we will take the matter field to be a
massless, minimally coupled scalar field, for which the stress tensor
is 
\begin{equation}
T_{\mu \nu} = \phi_{,\mu} \phi_{,\nu} - \frac{1}{2} g_{\mu\nu}\,
                \phi^{,\rho} \phi_{,\rho} \,,
\end{equation}
and the Ricci tensor is
\begin{equation}
R_{\mu \nu} = 8 \pi \,\phi_{,\mu} \phi_{,\nu} \,.
\end{equation}
This will now be interpreted as a normal ordered operator, so the correlation 
function $C_{\mu \nu \alpha \beta}(x,x')$ can be expressed as a sum of a
fully normal ordered term, a cross term, and a vacuum term:
\begin{equation}
 C^{\mu \nu \alpha \beta}(x,x') = C^{\mu \nu \alpha \beta}(x,x')_{NO}
+ C^{\mu \nu \alpha \beta}(x,x')_{cross} + C^{\mu \nu \alpha \beta}(x,x')_{vac}
  \, .
\end{equation}
Each of these terms can be found explicitly from Eqs.~(\ref{eq:S0}), 
(\ref{eq:S1}), and (\ref{eq:S2}) to be
\begin{equation}
C^{\mu \nu \alpha \beta}(x,x')_{NO} = 
64 \pi^2 \langle :\partial^\mu \phi(x)\, \partial^\nu \phi(x)\,
\partial'^\alpha \phi(x') \, \partial'^\beta \phi(x'): \rangle
 \,,     \label{eq:CNO}
\end{equation}
\begin{eqnarray}
C^{\mu \nu \alpha \beta}(x,x')_{cross} &=&
64 \pi^2 \left(\partial^\mu \partial'^\alpha D \; 
\partial^\nu \partial'^\beta D_0
+ \partial^\nu \partial'^\alpha D \; \partial^\mu \partial'^\beta D_0
    \right. \nonumber \\
&+& \left. \partial^\nu \partial'^\beta D \;\partial^\mu \partial'^\alpha D_0 
+  \partial^\mu \partial'^\beta D \; \partial^\nu \partial'^\alpha\ D_0 \right)
\, ,
\end{eqnarray}
and 
\begin{equation}
C^{\mu \nu \alpha \beta}(x,x')_{vac} = 64 \pi^2 \left(
\partial^\mu \partial'^\alpha D_0 \; \partial^\nu \partial'^\beta D_0
+ \partial^\nu \partial'^\alpha D_0 \; 
                 \partial^\mu \partial'^\beta D_0 \right) \,.
\end{equation}
Here $\partial^\mu$ denotes differentiation at point $x$ and 
$\partial'^\alpha$ that at $x'$. The vacuum two-point function is
\begin{equation}
 D_0 =  \langle \phi(x) \phi(x') \rangle_0 = 
\frac{1}{4 \pi^2\, (x-x')^2} \, ,
\end{equation}
and
\begin{equation}
 D =  \langle :\!\phi(x) \phi(x')\!: \rangle
\end{equation}
is the renormalized two-point function in the chosen state.

The central quantity which we calculate is the variance of the expansion
parameter, $\langle (\Delta \theta)^2 \rangle$. However, it will be useful 
to relate it to more directly observable quantities. Two possibilities are
characteristic angle by which an images will be blurred by metric fluctuations,
$\Delta \varphi$, and the fractional fluctuations in the apparent 
luminosity of the source,
$\Delta L/L$. In Appendix A, we give a heuristic treatment of the relation
between $\Delta \varphi$ and $\Delta \theta$, and argue that we should expect
a blurring angle of the order of $\Delta \varphi = s \,\Delta \theta/2$,
where $s$ is the distance to the source. Our treatment is heuristic because
we actually analyze the situation of a source in flat spacetime whose position
fluctuates. We conjecture that a careful treatment of the effects of a
fluctuating spacetime geometry will give a result of the same order of
magnitude. Fluctuations of $\theta$ are also related to fluctuations in the
brightness, or apparent luminosity of the source, as is discussed in Appendix B.
In general, $\Delta L/L$ is given in terms of an integral of the expansion 
correlation function, however, in some cases, it is of the same order as
$\Delta \varphi$.

\subsection{Averaging over a Bundle of Geodesics}
\label{sec:ave}

In order for the contributions of the cross and vacuum terms to be finite,
it is necessary to average over a spacetime volume. This volume is the interior
of the world tube defined by the bundle of geodesics. Thus in Eq.~(\ref{eq:var}) ,
we replace the integrations on the affine parameter by four-dimensional
spacetime integrations and write
\begin{equation}
\langle (\Delta \theta)^2 \rangle =
\int d^4 x \int d^4 x'' \, f(x) \, f(x') \,
 C_{\mu \nu \alpha \beta}(x,x')\, k^\mu (x) k^\nu (x)
\, k^\alpha(x') k^\beta(x') \,.  \label{eq:STT}
\end{equation}
where $f(x)$ is the function which describes the shape of the world tube.

It is convenient to express this averaging in null coordinates. Let the
unperturbed rays propagate in the $x$-direction, and let $u=t-x$ and
$v=t+x$ be the null coordinates. We can take the affine parameter to be the
$v$- coordinate, and then introduce additional averaging in the $u$, $y$,
and $z$ directions. We will take the averaging functions to be Lorenztian
or Gaussian functions in each of these directions. Let
\begin{equation}
g_L(u,\alpha) = \frac{\alpha}{\pi (u^2 + \alpha^2)} \, ,
\end{equation}
and
\begin{equation}
g_G(u,\alpha) = \frac{1}{\sqrt{\pi}\,\alpha} \, {\rm e}^{-u^2/\alpha^2} 
\end{equation}
so that
\begin{equation}
\int_{-\infty}^{\infty} g_L(u,\alpha) \, du = 
\int_{-\infty}^{\infty} g_G(u,\alpha) \, du = 1 \,.
\end{equation}
Equation~(\ref{eq:STT})  involves an averaging in six variables, 
$u$, $u'$, $y$, $y'$,
$z$, and $z'$, as well as integrations over $v$ and $v'$. However, in the
cases that will be of interest, the correlation function depends only upon
the differences $\Delta u = u -u'$, etc. We can exploit the property of
Lorenztian and  Gaussian functions that averaging in $u$ and $u'$
separately is equivalent to an average on $\Delta u$. In particular,
\begin{equation}
\int_{-\infty}^{\infty} du \int_{-\infty}^{\infty} du' \, g_L(u,\alpha) \,
g_L(u',\alpha) \, F(u-u') 
= \int_{-\infty}^{\infty} d \Delta u \, g_L(\Delta u,2\alpha)\, F(\Delta u) \,,
\end{equation}
and 
\begin{equation}
\int_{-\infty}^{\infty} du \int_{-\infty}^{\infty} du' \, g_G(u,\alpha) \,
g_G(u',\alpha) \, F(u-u') 
= \int_{-\infty}^{\infty} d \Delta u \, 
g_G(\Delta u,\sqrt{2} \alpha)\, F(\Delta u) \,.
\end{equation}
Then we need only average over the differences $\Delta u$, $\Delta y$, and
$\Delta z$. It will be convenient to 
let the averaging function be a Lorenztian of width $a$ in 
$\Delta u$ and Gaussians of width $b$ in each of $\Delta y$ and $\Delta z$,
so
\begin{equation}
f(x) = g_L(\Delta u,a)\,g_G(\Delta y,b)\,g_G(\Delta z,b) \,.
\end{equation}
The integrations on $v$ and $v'$ already amount to averaging, so we do not 
need to introduce any additional smearing in these variables. If the
integrand is a function of the difference $\Delta v = v-v'$ then we may use the
relation
\begin{equation}
\int_0^{v_0} dv \int_0^{v_0} dv' \, F(v-v') = 
\int_0^{v_0} d\Delta v \,(v_0-\Delta v)\,F(\Delta v) +
\int_{-v_0}^0 d\Delta v \,(v_0+\Delta v)\,F(\Delta v) \, . \label{eq:int_form1}
\end{equation}
If $F(\Delta v)$ is an even function, this relation becomes
\begin{equation}
\int_0^{v_0} dv \int_0^{v_0} dv' \, F(v-v') = 
2 \int_0^{v_0} d\Delta v \,(v_0-\Delta v)\,F(\Delta v) \, . \label{eq:int_form2}
\end{equation}

Henceforth, we will slightly change the notation, dropping the "$\Delta$"s,
and write $v$, $u$, $y$, and $z$ for $\Delta v$, $\Delta u$, $\Delta y$, and 
$\Delta z$, respectively. Now our expression for the averaged variance in
the expansion becomes
\begin{eqnarray}
& & \langle (\Delta \theta)^2 \rangle = 
\left[\int_0^{v_0} d v \,(v_0- v)\, + \int_{-v_0}^0 d v \,(v_0+ v)\right] 
\int_{-\infty}^{\infty} du \,g_L(u,a)\, \times  \nonumber \\
  & & \int_{-\infty}^{\infty} dy\, 
\int_{-\infty}^{\infty} dz\, \,g_G(y,b)\,g_G(z,b) \,
 C_{\mu \nu \alpha \beta}(x,x')\, 
k^\mu (x) k^\nu (x)\, k^\alpha(x') k^\beta(x') \, . \label{eq:var_av}
\end{eqnarray}

\subsection{Analysis of the Cross and Vacuum Terms}
\label{sec:cross_vac}

Here we wish to look at the state-independent and potentially singular
factors in the cross and the vacuum terms.
Let
\begin{equation}
\sigma = \frac{1}{2} (x-x')^2 =  \frac{1}{2} \Delta x^\mu \,\Delta x_\mu \,.
\end{equation}
Then the vacuum two-point function is
\begin{equation}
D_0(\sigma) = \frac{1}{8 \pi^2\, \sigma} \,.
\end{equation}
Further let
\begin{equation}
B \equiv k_\mu k_\alpha \,\partial^\mu \partial'^\alpha D_0
  = -(k_\mu k^\mu)D'_0 - (k_\mu \Delta x^\mu)^2 D''_0 \,.
\end{equation}
Here $D'_0$ and $ D''_0$ are derivatives of $D_0$ with respect to $\sigma$.
The contribution of the vacuum term to 
$C_{\mu \nu \alpha \beta}\, k^\mu  k^\nu \, k^\alpha k^\beta$, the integrand 
in Eq.~(\ref{eq:var_av}), is
\begin{equation}
C_{\rm vac}^{\mu \nu \alpha \beta}\, k_\mu  k_\nu \, k_\alpha k_\beta
= 128 \pi^2\, B^2 \,. \label{eq:C_vac}
\end{equation}
Similarly, the contribution of the cross term is 
\begin{equation}
C_{\rm cross}^{\mu \nu \alpha \beta}\, k_\mu  k_\nu \, k_\alpha k_\beta
= 256 \pi^2 \, k_\mu k_\alpha \,(\partial^\mu \partial'^\alpha D) \; B \,.
  \label{eq:C_cross}
\end{equation}

The contributions of these terms must be defined by averaging, as in
Eq.~(\ref{eq:STT}). 
However, we can take the tangent vector $k^\mu$ to be null, even
under the averaging, so all of the rays in the bundle over which we average 
have the same direction. Then $k^\mu k_\mu =0$ and the first term in $B$
vanishes. We can take $k_\mu = -\delta_\mu^u$. then
\begin{equation}
B = - \frac{2 u^2}{\pi^2 (r^2 -u v)^3}\,,
\end{equation}
where $r^2 = y^2+z^2$.
The vacuum contribution to $\langle (\Delta \theta)^2 \rangle$ becomes
\begin{eqnarray}
& & \langle (\Delta \theta)^2 \rangle_{vac} = \frac{1024}{\pi^2}\,
\int_0^{v_0} d v \,(v_0- v)\, \times \nonumber \\
& & \int_{-\infty}^{\infty} du \,\int_{-\infty}^{\infty} dy\,
\int_{-\infty}^{\infty} dz\, g_L(u,a)\,g_G(y,b)\,
g_G(z,b) \,
\left[\frac{u^4}{\pi^2 (r^2 -u v)^6} \right] \,.
\end{eqnarray}
Note that once the integration on $u$ is performed, the integrand is an even 
function of $v$, so we can use Eq.~(\ref{eq:int_form2}). 

Let us first consider the $u$-integration. The integrand has simple poles at
$u= \pm i a$, coming from the sampling function $g_L(u,a)$. However, it also
has a sixth-order pole at $u=r^2/v$, which reflects the singular nature of the 
vacuum term. As is well-known, contour integrals around higher order poles 
may be evaluated by integration by parts. Here the answer must be a real
quantity, so we can write the integral on $u$ as the average of two integrals,
one over a contour closed in the upper-half plane and going above the 
sixth-order pole on the real axis, and a second over a contour closed in
the lower-half plane and going below this pole. The result is finite and
just the real part of the residue of the pole at either $u=ia$ or $u=-ia$,
\begin{equation}
\int_{-\infty}^{\infty} du \,g_L(u,a)\,
\left[\frac{u^4}{ (r^2 -u v)^6} \right] =
{\rm Re}\left[\frac{a^4}{(r^2 -i a v)^6} \right]\,.
\end{equation}
It is convenient to perform the $v$-integration next, which yields
\begin{equation}
\int_0^{v_0} d v \,(v_0- v)\, {\rm Re}\left[\frac{a^4}{(r^2 -i a v)^6} \right]
= \frac{a^4 v_0^2 (a^6 v_0^6 +4 a^4 v_0^4 r^4 + 5 a^2 v_0^2 r^8 +10 r^{12})}
{20 r^8 (a^2 v_0^2 + r^4)^4} \,.   \label{eq:v_int}
\end{equation}
In the limit that $v_0 \gg r^2/a$, this approaches $a^2/(20 r^8)$. Both the exact
form, Eq.~(\ref{eq:v_int}), and the large $v_0$ limit have the same $r^{-8}$
behavior for small $r$. In either case, a further averaging in the transverse
directions is needed to produce a finite result. 

In the large $v_0$ limit, we can write
\begin{equation}
\langle (\Delta \theta)^2 \rangle_{vac} = \frac{256}{5 \pi^2} \, a^2\,
\left\langle \frac{1}{r^8} \right\rangle \,.
\end{equation}
The evaluation of quantities such as $\langle {1}/{r^8} \rangle$ is 
discussed in Appendix C, where it is argued that 
\begin{equation}
\left\langle \frac{1}{r^8} \right\rangle = \frac{A_8}{ b^8}\, ,
\end{equation}
where $A_8$ has at most a logarithmic dependence on $b$.
This leads to our final 
result for the contribution of the vacuum term in the large $v_0$ limit,
\begin{equation}
\langle (\Delta \theta)^2 \rangle_{vac} = \ell_P^4\,
\frac{256\, A_8\, a^2}{5 \pi^2 b^8} \,, \label{eq:vac_theta}
\end{equation}
where $\ell_P$ is the Planck length.

Note that $v_0 = 2\,s$, where $s$ is the spatial separation between the
source and detector, and that the large $v_0$ limit now means $v_0 \gg b^2/a$.
The fact that $\langle (\Delta \theta)^2 \rangle_{vac}$ approaches a finite
value in this limit indicates that the focusing fluctuations in
the Minkowski vacuum do not accumulate, but they can be non-zero. This 
raises the question as to whether they lead to potentially observable effects.
The expected angular blurring or fractional luminosity fluctuations of an 
image at a distance $s$ are  given by Eqs.~(\ref{eq:delphi}) and
(\ref{eq:dellum}), respectively, and in this case will be
\begin{equation}
2 \Delta \varphi = 
\left({\Delta L\over L}\right)_{rms} = 0.1 \,A_8\, \ell_P^2 \, \frac{a\, s}{b^4} \approx
10^{-8}\,A_8\, \left(\frac{a}{b}\right)\, 
\left(\frac{10^{-10} {\rm cm}}{b}\right)^3 \,
\left(\frac{s}{10^{28} {\rm cm}}\right) \, .   \label{eq:vac_phi}
\end{equation}
The smallest reasonable averaging scale would be of the order of the wavelength
of the photons from the source. Even in this case, the effect predicted by
Eq.~(\ref{eq:vac_phi}) is too small to have been seen in any measurements 
that have yet been performed.

It is also of interest to note that the asymptotic value of 
$\langle (\Delta \theta)^2 \rangle_{vac}$ can be made arbitrarily small
by making the temporal averaging scale $a$ small, although the value
of $v_0$ required to attain this asymptotic value grows. This behavior can 
be traced to the factor of $(k_\mu \Delta x^\mu)^4$ in Eq.~(\ref{eq:C_vac}).
This factor in turn contributes a factor of $a^4$ to 
$\langle (\Delta \theta)^2 \rangle_{vac}$, and turns what would otherwise be
an $a^{-2}$ dependence into $a^{2}$ dependence. A general product of
stress tensor operators needs to be averaged in time to be finite, but
the particular combination that contribute to 
$\langle (\Delta \theta)^2 \rangle_{vac}$ remains finite even in the limit
that the time averaging scale goes to zero.

Now let us turn to the cross term, which is determined by 
Eq.~(\ref{eq:C_cross}). This term also requires averaging to render it finite.
However, it is state-dependent, so we cannot calculate it explicitly
without specifying the function $D$. However, we can look at the result
of averaging the state-independent factor $B$, and then assuming that the
state-dependent factor is approximately independent of $v$. This leads to
\begin{equation}
\langle (\Delta \theta)^2 \rangle_{cross} = 
512 \pi^2 \, k_\mu k_\alpha \,(\partial^\mu \partial'^\alpha D) \;
\int_0^{v_0} d v \,(v_0- v)\, \langle B \rangle \,.
\end{equation}
We may evaluate the integral of $\langle B \rangle$ in the same manner
as used for the vacuum term:
\begin{equation}
\langle B \rangle = \int d^2x \, g(r)\,\int_{-\infty}^{\infty} du \,
g_L(u,a)\, B = -\frac{2 a^2}{\pi^2}\, \int d^2x \, g(r)\,
\frac{r^2 (3 a^2 v^2 -r^4)}{( a^2 v^2 +r^4)^3} \,.
\end{equation}
Then 
\begin{equation}
\int_0^{v_0} d v \,(v_0- v)\,\langle B \rangle \,=
\frac{a^2}{\pi^2} \int d^2x \, g(r)\, \frac{a^2 v_0^2}{(a^2 v_0^2 +r^4)} \,.
\end{equation}
Note that this expression, and hence the contribution of the cross term,
vanishes in the limit that $a \rightarrow 0$ for fixed $v_0$. However,
if we let $v_0 \rightarrow \infty$ for fixed, nonzero $a$, then we find
\begin{equation}
\int_0^{v_0} d v \,(v_0- v)\,\langle B \rangle =
\frac{1}{\pi^2} \, \left\langle \frac{1}{r^2} \right\rangle =
\frac{A_2}{\pi^2\, b^2} \,, \label{eq:cross}
\end{equation}
where $A_2$ has at most a logarithmic dependence on $b$.
 As expected, the cross term is singular in the limit that
$b \rightarrow 0$, but less so than the vacuum term.

As was the case for the vacuum term, $\langle (\Delta \theta)^2 \rangle_{cross}$
vanishes in the limit that $a \rightarrow 0$. In both cases, this is due to the 
presence of the factor of  $(k_\mu \Delta x^\mu)^2$ in $B$. Note that if we were 
to take first the limit $a \rightarrow 0$ and then the limit $b \rightarrow 0$,
we would have a localized ray for which $\langle (\Delta \theta)^2 \rangle$
is given entirely by the fully normal ordered term. 

Here we have used a specific choice of averaging. One topic for future
research will be to generalize this choice. In particular, it will be
of interest to write covariant expressions for $B$ and $B^2$ as derivatives
of logarithmic functions, and then integrate over a four-dimensional
spacetime region.

\subsection{The Physical Meaning of Smearing Scales $a$ and $b$}

We have defined the singular terms of the Ricci tensor by averaging over a 
world tube containing the bundle of geodesics.   The shape of this world 
tube is determined by its spatial width $b$, which describes the transverse 
distance from the central geodesic, and  its temporal width $a$.      

As far as the temporal smearing is concerned, it is worth noting that the 
parameter $a$ only appears in the pure vacuum term, and does not affect
 results obtained for state-dependent terms (the cross terms and the fully 
normal-ordered term).   If one assumes that pure vacuum effects in Minkowski 
spacetime are unphysical, or least unobservable, a discussion of the meaning 
of $a$ is rendered superfluous.  However, if we remain open to the possibility 
that the pure vacuum effects possess physical significance, we should attempt 
to connect $a$ with some concrete length scale in the problem.  There are a 
few reasonable options to chose from.   Certainly the following come to mind:  
the wavelength of the source, the coherence length of the source, and the 
duration of time over which the light source is collected.  
The latter, in at least one respect, seems the least appealing.   
It would suggest that the vacuum effect could be made arbitrarily large 
by leaving the measuring device on for an arbitrarily long time 
(noting the power dependence in Eq.~(\ref{eq:vac_phi}).   
If we are concerned with finding a upper bound when the measuring process 
is arbitrarily extended in time, it seems more likely that the coherence 
length of the light source, or even the smallest scale involved, 
the wavelength, should be related to the temporal smearing.   
The argument here is that the photon extends off the lightcone, but 
certainly no more than the temporal size of its wave packet, and no less 
than the resolution provided by its characteristic wavelength.

On the other hand, it seems most natural to associate the width of the bundle 
with the parameters of the experimental set-up, determined by both conditions 
at the source and at the observer.   That is, it seems we can associate the 
width of the bundle $b$  with the two quantities:  the aperture of the device 
used to collect the light rays, and the size of the emitting source.   
If both could be fixed to the same size, we would effectively enclose the 
bundle of geodesics within a cylinder extending along the entire flight path.  
Under such a circumstance it seems safe to say that $b$ would simply be the 
radius of the cylinder.  In the case of  point source emissions, $b$ might 
be sensibly set to the average width of the bundle along the flight path, 
i.e. one half of the aperture's diameter.   For a larger source,  $b$ should 
perhaps an average between the diameter of the emitting source and the 
diameter of the aperture.

\section{A Thermal State}
\label{sec:thermal}

\subsection{Fully Normal Ordered Term in the Low Temperature Limit}
\label{sec:thermal_FNO}

Now we turn to the case of a thermal state, so the fluctuations of the expansion
have their origin in thermal fluctuations of the scalar field stress tensor.
In this subsection, we consider the limit of a localized ray in the sense
discussed above. The limit in which $a$ and $b$ are small is equivalent to a 
low temperature limit because the thermal wavelength is large compared to
these scales. In this limit, we need to consider  
only the effect of the fully normal ordered term, 
so we must calculate the right-hand side of Eq.~(\ref{eq:CNO}).
 There is a useful identity which can be used for this purpose:
\begin{equation}
\langle :\!\phi_1 \phi_2 \phi_3 \phi_4\!: \rangle =
\langle :\!\phi_1 \phi_2\!: \rangle \langle :\!\phi_3 \phi_4\!: \rangle +
\langle :\!\phi_1 \phi_3\!: \rangle \langle :\!\phi_2 \phi_4\!: \rangle +
\langle :\!\phi_1 \phi_4\!: \rangle \langle :\!\phi_2 \phi_3\!: \rangle \,,
                                                \label{eq:ident}
\end{equation}
where the expectation values are taken in a thermal state. One proof of this 
identity is based upon the Hawking effect~\cite{Hawking}. First, one may use
Wick's theorem to prove Eq.~(\ref{eq:ident}) for a generalized vacuum 
state~\cite{Kuo}. This is any state $|\psi \rangle$ for which there exists 
a decomposition of the field operators into positive and negative frequency
parts, 
\begin{equation}
\phi = \phi^+ + \phi^- \, , \label{eq:posneg}
\end{equation}
such that $\phi^+ |\psi \rangle =
\langle \psi | \phi^- = 0$. However, the formation of a black hole by 
gravitational collapse transforms the in-vacuum state into a thermal state 
at late times~\cite{Hawking}. As a consequence, Eq.~(\ref{eq:ident}) must also be valid for
thermal states.

It is also possible to give a more direct proof of this identity.   Recall
that the thermal average of an operator $A$ can be expressed as
\begin{equation}
\langle A \rangle_\beta = \frac{{\rm Tr} (\rho \, A)}{{\rm Tr}(\rho)} \, ,
\end{equation}
where $\rho$ is the density operator. Let us now express each of the
$\phi_i$ operators as a sum of positive and negative frequency components, as
in Eq.~(\ref{eq:posneg}). Recall that normal ordering means positive
frequency terms are to the right of all negative frequency terms. Because
$\rho$ has a Fock space representation as a sum of terms of the form
$|n\rangle \langle n|$, where $|n\rangle$ is a number eigenstate, only
terms with equal numbers of positive frequency and  negative frequency
factors can have nonzero expectation values. Thus, we can write
\begin{eqnarray}
\langle :\!\phi_1 \phi_2 \phi_3 \phi_4\!: \rangle &=&
\langle \phi_1^- \phi_2^- \phi_3^+ \phi_4^+ \rangle +
\langle \phi_1^- \phi_3^- \phi_2^+ \phi_4^+ \rangle +
\langle \phi_1^- \phi_4^- \phi_2^+ \phi_3^+ \rangle \nonumber \\
&+& \langle \phi_2^- \phi_3^- \phi_1^+ \phi_4^+ \rangle +
\langle \phi_2^- \phi_4^- \phi_1^+ \phi_3^+ \rangle +
\langle \phi_3^- \phi_4^- \phi_1^+ \phi_2^+ \rangle  \, .
\end{eqnarray}
Next we can use the fact that in a thermal state 
\begin{equation}
\langle a^\dagger_i a^\dagger_j a_k a_l \rangle =
\langle a^\dagger_i a_l \rangle \langle a^\dagger_j a_k \rangle +
\langle a^\dagger_i a_k \rangle \langle a^\dagger_j a_l \rangle
\end{equation}
to write
\begin{equation}
\langle \phi_1^- \phi_2^- \phi_3^+ \phi_4^+ \rangle =
\langle \phi_1^- \phi_4^+ \rangle \langle \phi_2^- \phi_3^+ \rangle +
\langle \phi_1^- \phi_3^+ \rangle \langle \phi_2^- \phi_4^+ \rangle \, ,
                                                   \label{eq:quarquad}
\end{equation}
and similar relations. We can now express 
$\langle :\!\phi_1 \phi_2 \phi_3 \phi_4\!: \rangle$ as a sum of products of
the form of that on the right-hand side of  Eq.~(\ref{eq:quarquad}).
By use of the fact that for a thermal state
\begin{equation}
\langle :\phi_1 \phi_2 : \rangle = \langle \phi_1^- \phi_2^+ \rangle +
\langle \phi_2^- \phi_1^+ \rangle \, ,
\end{equation}
this sum can be rewritten to complete the proof of Eq.~(\ref{eq:ident}).

By use of this identity, the fully normal ordered term, Eq.~(\ref{eq:CNO}),
can be expressed as
\begin{equation}
C^{\mu \nu \alpha \gamma}(x,x')_{NO} = 64 \pi^2 \left(
\partial^\mu \partial'^\alpha D_\beta \; \partial^\nu \partial'^\gamma D_\beta
+ \partial^\nu \partial'^\alpha D_\beta \; 
\partial^\mu \partial'^\gamma D_\beta \right) \,,
\end{equation}
where 
\begin{equation}
 D_\beta =  \langle :\!\phi(x) \phi(x')\!: \rangle_\beta = 
{\sum_{n=-\infty}^{\infty}}' \frac{1}{4 \pi^2[|\Delta {\bf x}|^2 - 
                                      (\Delta t +i n \beta)^2]} \, ,
\end{equation}
is the renormalized thermal two-point function. Here $\beta =1/T$,
where $T$ is the temperature, 
$\Delta {\bf x} = {\bf x} - {\bf x'}$, $ \Delta t = t -t'$, and the prime
on the summation denotes that the $n=0$ term is omitted.

Recall that here we are discussing a limit of localized rays.
Let the  unperturbed path of the bundle of geodesics be in the $+x$-direction,
so that $\Delta x = \Delta t$ and $\Delta y = \Delta z =0$. We can take
the coordinate time in a given frame of reference to be the affine parameter,
so that $\lambda = t$, $\lambda' = t'$ and $k^\mu = (1,1,0,0)$ in 
Eq.~(\ref{eq:var}). After some calculation, the integrand may be expressed as 
\begin{equation}
C^{\mu \nu \alpha \gamma}\, v_\mu  v_\nu \, v_\alpha v_\gamma =
C_{NO}^{\mu \nu \alpha \gamma}\, v_\mu  v_\nu \, v_\alpha v_\gamma =
\frac{2^{11}}{\pi^2\, \beta^8}\> f(k) \, ,
\end{equation}
where
\begin{eqnarray}
f(k) &=& \left[ \sum_{n=1}^\infty \frac{(n^2 - 3k^2)}{(n^2+k^2)^3} \right]^2
                                              \nonumber \\
&=& \frac{1}{4 k^8}\, \left[ \pi^3 k^3 \,\coth (\pi k) \, {\rm csch}^2 (\pi k) 
+ \pi^2 k^2 \, {\rm csch}^2 (\pi k) + \pi k \,\coth (\pi k) \, - 3 \right]^2 
\end{eqnarray}
and $k = 2 \Delta t/\beta$. Now we can write the variance of the expansion as
\begin{equation}
\langle (\Delta \theta)^2 \rangle  =
\int_0^s d t \, \int_0^s d t' \; 
C_{NO}^{\mu \nu \alpha \gamma}\, k_\mu  k_\nu \, k_\alpha k_\gamma =
\frac{2^{10}}{\pi^2\, \beta^6}\> \int_0^\kappa d k (\kappa -k) f(k) \,.
                                                 \label{eq:var2}
\end{equation}
Here $s$ is the separation of the source and detector (the flight time),
$\kappa = 2 s/\beta$, and we have used Eq.~(\ref{eq:int_form2}).
The function $f(k)$ is finite at $k = 0$,
$f(0) = \pi^8/8100$ and vanishes rapidly for large $k$:
\begin{equation}
f(k) \sim \frac{\pi^2}{4 k^6} \, , \quad k \rightarrow \infty \, .
\end{equation}
As a consequence, both $\int_0^\infty d k \, f(k)$ and 
$\int_0^\infty d k \,k \, f(k)$ are finite. In this case, the large $\kappa$
limit of  Eq.~(\ref{eq:var2}) is
\begin{equation}
\langle (\Delta \theta)^2 \rangle = 
\frac{2^{10}\, c_0}{\pi^2\, \beta^6}\, \kappa \, ,
\end{equation}
where
\begin{equation}
c_0 = \int_0^\infty d k \, f(k) \approx 0.3468 \,.
\end{equation}

We can now write the root-mean-squared fluctuation in the expansion as
\begin{equation}
\Delta \theta_{rms}= \frac{128 \sqrt{c_0}}{\pi} \, \ell_P^2 \, \sqrt{s\, T^7} \, .
                                                \label{eq:del_theta}
\end{equation}
 The fractional luminosity fluctuations of the image
of a point source viewed at a distance $s$ through a bath of scalar radiation
at temperature $T$ becomes, for localized rays,
\begin{equation}
\left({\Delta L \over L}\right)_{rms} = 
\frac{64 \sqrt{c_0}}{\sqrt{3}\,\pi} \, \ell_P^2 \, \sqrt{s^3\, T^7} \, .
                                    \label{eq:del_phi}
\end{equation}

Equation~(\ref{eq:del_phi}) can be expressed as
\begin{equation}
\left({\Delta L \over L}\right)_{rms} = 0.02\, 
\left(\frac{s}{10^{28} {\rm cm}} \right)^\frac{3}{2}
\, \left(\frac{T}{10^6 K} \right)^\frac{7}{2}  =
10^{-3}\, \left(\frac{s}{10^{6} {\rm km}} \right)^\frac{3}{2}
\, \left(\frac{T}{1 {\rm GeV}} \right)^\frac{7}{2} \,. \label{eq:estimates}
\end{equation}
Thus a source seen at a cosmological distance, $10^{28} {\rm cm} \approx
3 {\rm Gpc}$ through a thermal bath at a temperature of $10^6 K$ could
in principle show large luminosity fluctuations if one could select rays 
localized on a scale less than the thermal wavelength $\beta$.
 Of course, there cannot be such
a bath in our present universe, as its energy density would have long dominated
the expansion of the universe. Note that $10^6 {\rm km}$ is the 
Schwarzschild radius of a black hole with a mass of $3 \times 10^5 M_\odot$,
in the mass range of the supermassive black holes which may be in
the centers of many galaxies, and possible linked to the phenomenon of gamma
ray bursts. In at least some cases~\cite{Biller}, these bursts have produced 
gamma rays with energies well in excess of $1 {\rm GeV}$. Although the details
of the size and nature of the emitting regions is unclear, this raises the
possibility that within the burst sources, there are quantum metric
fluctuations large enough to achieve $\Delta L/ L \approx 1$ for a bundle
of rays localized on a scale small compared to thermal wavelength $\beta$.
Any realistic observation is likely to average over a much larger scale, so we
now turn to the opposite limit.

\subsection{High Temperature Limit}
\label{sec:hightem}

In the limit in which the averaging scales $a$ and $b$ are large compared 
to the thermal wavelength, we need to adopt a different approach. Now the
effect of the averaging cannot be ignored, even in the fully normal-ordered
term. It is easiest to include the effects of the vacuum, cross, and 
fully normal-ordered terms all together by using the full thermal
Green's function,
\begin{equation}
 D_{F \beta} =  \langle :\!\phi(x) \phi(x')\!: \rangle_\beta = 
{\sum_{n=-\infty}^{\infty}} \frac{1}{4 \pi^2[|\Delta {\bf x}|^2 - 
                                      (\Delta t +i n \beta)^2]} \, .
\end{equation}
If we impose the same type of averaging as used in Sect.~\ref{sec:cross_vac}, 
then we can write
\begin{eqnarray}
\!\!\!\!\!\!\!\!\!\!\!\!& & \langle (\Delta \theta)^2 \rangle = 
      \nonumber \\
\!\!\!\!\!\!\!\!\!\!\!\!& & \frac{512}{\pi^2} \,
\left[\int_0^{v_0} d v \,(v_0- v)\, + \int_{-v_0}^0 d v \,(v_0+ v)\right] 
\int_{-\infty}^{\infty} du \,g_L(\Delta u,a)\, 
\int d^2x \, g(r)\, (\partial_v^2 D_{F \beta})^2 \,.
\end{eqnarray}
In the high temperature (small $\beta$) limit, we can replace the sum
on $n$ in $ D_{F \beta}$ by an integral. Let $\xi = \beta n$, and let
\begin{equation}
\sum_{n=-\infty}^{\infty} \rightarrow 
\frac{1}{\beta} \int_{-\infty}^{\infty} d \xi \,.
\end{equation}
Then we can write
\begin{equation}
\partial_v^2 D_{F \beta} = \frac{16}{\pi^2 \beta}\,
\int_{-\infty}^{\infty} d\xi\, 
\frac{(i\xi+u)^2}{[(v-u)^2+4r^2 -(2i\xi +v+u)^2]^3}
= -\frac{1}{2\pi \beta}\, \frac{2r^2 -(v-u)^2}{[(v-u)^2 +4r^2]^\frac{5}{2}} \,. 
\end{equation}
This leads to
\begin{eqnarray}
& & \langle (\Delta \theta)^2 \rangle =
\frac{32}{\beta^2}\, \int_{-\infty}^{\infty} du \,g_L(\Delta u,a)\,
\int d^2x \, g(r)\, \times   \nonumber \\
& & \left[\int_0^{v_0} d v \,(v_0- v)\, + \int_{-v_0}^0 d v \,(v_0+ v)\right]
\; \frac{[2r^2 -(v-u)^2]^2}{[(v-u)^2 +4r^2]^5} \, .
\end{eqnarray}
In the limit that $v_0$ is large compared to either $a$ or $b$, we can
write this as
\begin{equation}
\langle (\Delta \theta)^2 \rangle =
\frac{32\,v_0}{\beta^2}\, \int_{-\infty}^{\infty} du \,g_L(\Delta u,a)\,
\int d^2x \, g(r)\,
\int_{-\infty}^{\infty} d v \,\frac{(2r^2 -v^2)^2}{(v^2 +4r^2)^5} 
= \frac{27 \pi \,v_0}{512 \beta^2}\, \left\langle \frac{1}{r^5} \right\rangle\,.
                                \label{eq:HiT1}
\end{equation}
Note that in this limit, the integrand of the $v$-integration becomes
independent of $u$, and hence the result is independent of $a$.

The final expression involves the transverse average of a singular
function, $1/r^5$. However, unlike the singular functions encountered
earlier in the vacuum and cross terms, this is simply an artifact of our 
approximations. The dominant contribution here is from the fully-normal
ordered term, and should be finite. The apparent singularity at $r=0$
arises because our approximation of replacing the sum by an integral
assumes that $\beta$ is small compared to all other length scales including
$r$. An exact calculation of the fully-normal ordered contribution to 
$\langle (\Delta \theta)^2 \rangle$ would, in the high temperature limit,
replace Eq.~(\ref{eq:HiT1}) by
\begin{equation}
\langle (\Delta \theta)^2 \rangle = \frac{27 \pi \,v_0}{512 \beta^2}\, 
\left\langle f(r) \right\rangle\,,    \label{eq:HiT2}
\end{equation}
where $f(r)$ is a function which falls as $r^{-5}$ for $r \agt \beta$, but
approaches a constant for $r \alt \beta$. We can estimate $\langle f(r) \rangle$
in the high temperature limit by simply cutting off the integration on $r$ at
a lower limit of order $\beta$. Thus,
\begin{equation}
\langle f(r) \rangle \approx \frac{2}{b^2}\, \int_{\beta}^{\infty}
\frac{d r}{r^4} \, {\rm e}^{-r^2/b^2} \approx \frac{2}{3 b^2 \beta^3} \,.
\end{equation}
This leads to an estimate for $\langle (\Delta \theta)^2 \rangle$ of
\begin{equation}
\langle (\Delta \theta)^2 \rangle = \frac{K^2}{b^2\, \beta^5}\, s \,,
\end{equation}
where $K$ is a constant of order one.

The corresponding fractional luminosity fluctuation is
\begin{equation}
\left({\Delta L \over L}\right)_{rms} = 
\frac{K\,s^\frac{3}{2}}{b \, \beta^\frac{5}{2}}\, \ell_P^2 \,.
                                    \label{eq:del_phi_high}
\end{equation}
Note that as compared to Eq.~(\ref{eq:del_phi}), the effect is now 
suppressed by a factor of $\beta/b$. This is essentially the effect of classical 
averaging of the luminosity fluctuations; when one averages over a larger 
source, the fluctuations are suppressed. In Eq.~(\ref{eq:del_phi_high}),
we might take $b$ to be either of the order of the size of the source,
or of the average of the source size and the observer's aperture size.
In either case, the luminosity fluctuations will be strongly suppressed 
compared to the estimates given in Eq.~(\ref{eq:estimates}).

\section{Discussion}
\label{sec:final}

In this paper, we have treated the effects of the fluctuations of a massless,
minimally coupled scalar field stress tensor on a flat background. 
In particular,
we used the fluctuations in expansion of a congruence of null rays as 
an operational measure of the effects of stress tensor fluctuations on
the spacetime geometry. Although the cross and vacuum terms are potentially
singular parts of the stress tensor correlation function, we argued that
their contributions can be rendered finite by an integration by parts
procedure. In the limit of localized rays, 
they actually yield a zero contribution to the
variance of the expansion. This is in contrast to other 
situations~\cite{WKF01,WF01}, where the cross term often yields the dominant
contribution. We calculated an explicit formulas 
for the fluctuations in the expansion produced by a thermal bath. This
in turn led to a results, Eqs.~(\ref{eq:del_phi}) and (\ref{eq:del_phi_high}), 
for the fractional luminosity fluctuations
of the image of an object seen through a fluctuating spacetime filled
with a thermal bath. Although these effects are usually small, there are
conceivable astrophysical situations where $\Delta L/L$ would
become large if the rays could be sufficiently localized. 

The effects of the fully-normal ordered term are unambiguous, but those
of the vacuum and cross terms contain singular parts which need to be defined 
by a spacetime averaging over a bundle of rays. This may be done by the 
integration by parts procedure discussed in Appendix C. This procedure involves
some undetermined constants, some of which can be eliminated by a scaling 
argument. Unfortunately, the overall constant which determines the magnitude
of the vacuum contribution is not fixed. Further work is needed to
determine this constant uniquely. As noted above, one topic for future
research will be the use of covariant averaging.

Note that Eq.~(\ref{eq:del_theta}) describes the effects of the Ricci tensor 
fluctuations, not the expansion due to any mean Ricci curvature. Our treatment
has assumed a nearly flat background spacetime, More generally, the non-zero
mean value of the stress tensor will lead to mean Ricci curvature and hence
classical focusing.

Recall that we assumed that the $\theta^2$ term in the Raychaudhuri equation,
Eq.~(\ref{eq:ray}), can be neglected. As $\langle (\Delta \theta)^2 \rangle$
grows, this assumption will eventually break down. However, in many cases
$\langle (\Delta \theta)^2 \rangle$ remains small compared to the Ricci
tensor term for a very long time. We can give an explicit estimate of this
time for the case of the thermal bath. In this case, the 
root-mean-square fluctuations
in the Ricci tensor term are of order $|R_{\mu\nu} v^\mu v^\nu|_{rms}
\approx \ell_P^2\, T^4$. Thus $\langle (\Delta \theta)^2 \rangle \ll 
|R_{\mu\nu} v^\mu v^\nu|_{rms}$ so long as
\begin{equation}
s \ll \frac{\beta^3}{\ell_P^2} \,.
\end{equation}
For temperatures less than about $1\,{\rm GeV}$, this condition will
be satisfied for any $s$ less than the size of the observable universe.
 We also assumed that the shear and vorticity terms in the 
Raychaudhuri equation can be ignored. So long as the bundle of geodesics
is hypersurface orthogonal, the vorticity remains zero in all classical
spacetimes, so it is reasonable to expect that small quantum geometry
fluctuations will not generate any vorticity. The shear can grow in
response to non-zero Weyl curvature. Thus if the stress tensor fluctuations
cause the Weyl tensor to fluctuate, through the excitation of gravitational
wave modes, then eventually $\langle \sigma_{\mu\nu} \sigma^{\mu\nu} \rangle$
could become important. However, it seems reasonable to speculate that it will
not grow any more rapidly than does $\langle (\Delta \theta)^2 \rangle$.

In summary, we have illustrated how  expansion fluctuations
provide a concrete measure of the effects of passive quantum geometry
fluctuations.

\begin{acknowledgments}
We would like to thank Arvind Borde, Ken Olum, Tom Roman, Richard Woodard, and 
Chun-Hsien Wu for useful discussions.
  This work was supported in part by the National
Science Foundation under Grant PHY-0244898.
\end{acknowledgments}

\appendix
\section{}

In this Appendix, we will illustrate the geometric meaning of the expansion
$\theta$ in an explicit example~\cite{Lud}, and then use this example 
to infer a
relation between fluctuations of the expansion and the angular blurring
of an image. First consider a point source in flat spacetime which is
emitting null rays, whose tangent vector field is $v^\mu = dx^\mu/d\lambda$.
Choose a coordinate system in which the origin of spatial coordinates is at
the location of the point source, and take the affine parameter to be 
coordinate time in this frame:  $\lambda = t$. Then
\begin{equation}
v^\mu = (1, {\bf v}) \, ,
\end{equation}
where ${\bf v}$ is a unit 3-vector, which is also the unit position vector
in this coordinate system
\begin{equation}
{\bf v} = \hat{\bf x} = \frac{\bf x}{|{\bf x}|} \,.
\end{equation}

The expansion is defined as the trace of the tensor $\nabla^\nu \,v^\mu$ in
the subspace orthogonal to $v^\mu$:
\begin{equation}
\theta = \nabla^\nu \,v^\mu\,(g_{\mu\nu} + v_\mu v_\nu) \,.
\end{equation}
In our case, the tensor $\nabla^\nu \,v^\mu$ has only spatial components
\begin{equation}
\nabla^j \,v^i = \partial^j\, v^i 
= \frac{\delta^{ij}\, |{\bf x}|^2 -x^i x^j}{|{\bf x}|^3} \,.
\end{equation}
Now $\theta$ can be found to be
\begin{equation}
\theta = \partial^j\, v^i \,(\delta_{ij} + v_i v_j) = \frac{2}{s} \, ,
\end{equation}
where $s = |{\bf x}|$ is the distance from the source. If we repeat
this calculation for the case of timelike geodesics, using the 4-velocity
$u^\mu$ as the tangent vector, the result is
\begin{equation}
\theta = \frac{2\,\gamma\,v}{s} \, ,
\end{equation}
where $\gamma = 1/\sqrt{1-v^2}$. In both cases, $\theta \propto 1/s$,
so as the distance from the source increases and the rays become more
nearly parallel, the expansion decreases.

\begin{figure}
\begin{center}
\leavevmode\epsfysize=6cm\epsffile{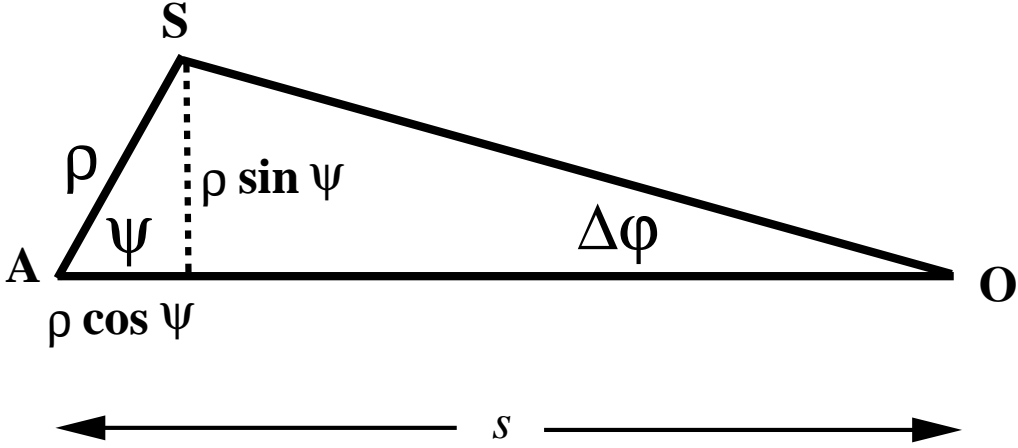}
\end{center}
\caption{A source $S$ at a mean distance $s$ from an observer $O$ is 
displaced in direction $\psi$ by a distance $\rho$ from its average 
position $A$. The effect is to shift the apparent
position of the source by an angle $\Delta \varphi$. }
\label{fig:source}
\end{figure}

Now let us relate the fluctuations of $\theta$ to fluctuations in the apparent 
angular position of the source. In our case, the $\theta$-fluctuations
are due to fluctuations in the spacetime geometry between the source and the
observer. However, one could achieve the same effect with a source whose 
position fluctuates in a fixed Minkowski spacetime. Consider the situation
illustrated in Fig.~\ref{fig:source}, where the source is displaced from 
its mean position
by a distance $\rho$ in a direction given by the angle $\psi$. For an observer
at a mean distance of $s$, the expansion changes by
\begin{equation}
\Delta \theta = -\frac{2}{s^2}\, \Delta s = \frac{2}{s^2}\,\rho \, \cos \psi \,,
\end{equation}
and the angular position of the source changes by approximately
\begin{equation}
\Delta \varphi = \frac{\rho \, \sin \psi}{s} \, ,
\end{equation}
provided that $\rho \ll s$. We now assume that $\rho$ and $\psi$ fluctuate
independently of one another and that all values of $\psi$ are equally
probable. Then
\begin{equation}
\langle \rho^2 \, \sin^2 \psi \rangle = 
\langle \rho^2 \rangle \, \langle \sin^2 \psi \rangle =
\frac{1}{2} \langle \rho^2 \rangle
\end{equation}
and 
\begin{equation}
\langle \rho^2 \, \cos^2 \psi \rangle = 
\langle \rho^2 \rangle \, \langle \cos^2 \psi \rangle =
\frac{1}{2} \langle \rho^2 \rangle \, .
\end{equation}
This would happen, for example, if $\rho$ does not fluctuate and $\psi$
fluctuates randomly. This allow us to relate the expansion fluctuations 
to the angular position fluctuations:
\begin{equation}
\langle (\Delta \varphi)^2 \rangle = 
\frac{1}{4}s^2\, \langle (\Delta \theta)^2 \rangle \,.
\end{equation}
Thus the effect of $\theta$-fluctuations is to blur the angular resolution
of the image of the source by an amount
\begin{equation}
\Delta \varphi_{\rm rms} = \frac{1}{2}\, s \, \Delta \theta_{\rm rms} \,.
                                            \label{eq:delphi}
\end{equation}
Strictly, we have treated only a source in flat spacetime whose position 
fluctuates. However, we offer this as a heuristic estimate of the
magnitude of angular blurring expected in a fluctuating spacetime.

\section{}

In this appendix, we show how the stochastic behavior of the expansion 
leads to small deviations in the density of geodesics.  In the case of 
light rays this translates into luminosity fluctuations.  The connection 
between luminosity fluctuations and expansion fluctuations can be seen as 
follows.   Let us write the expansion as the sum of a classical 
(deterministic) component $\theta_{c}$ and a stochastic component 
$\hat{\theta}$:
\begin{equation}
\theta(t) = \theta_{c}(t) + \hat{\theta}(t).
\end{equation}
Here $t$ is an affine parameter along the bundle of geodesics.  
The stochastic component can expresse as
\begin{equation}
\hat{\theta} = {d \log ({A}/A_{c}) \over dt}\, ,
\end{equation}
where $\hat{A}$ is the cross sectional area of a bundle.
Now we assume $ \hat{\theta}$ is a small stochastic perturbation.  This means
\begin{eqnarray}
{A} =  \hat{A}  + A_{c} \approx A_{c},
\end{eqnarray}
where $\hat{A}$ is the stochastic variation in the area (zero at $t=0$). 
Using the Taylor expansion of log(1+x) for small x, we find
\begin{eqnarray}
\hat{\theta}& \approx & {d \left(\hat{A}/A_{c}\right) \over dt},\\
\int_{t_{0}}^{t} dt' \; \theta(t') & \approx & { \hat{A} \over A_{c}} 
= {\Delta \rho \over \rho},
\end{eqnarray}
where $\rho$ is the density of geodesics.  Since the number of geodesics, $n$, 
is conserved (total energy flow), while the transverse area $A$ is distorted 
(density fluctuations),
\begin{eqnarray}
Intensity\;or\; Luminosity \;  & =&  {n \over A}.
\end{eqnarray}
The fractional luminosity fluctuation is therefore given by
\begin{eqnarray}
{ \Delta L \over L}   &= &{\Delta \rho \over \rho}\; \approx \; \int_{t_{0}}^{t} dt' \; \theta(t').
\end{eqnarray}
Now the variance of the fractional luminosity fluctuations can be computed as
\begin{equation}
\left\langle\;\left(\Delta L \over L\right)^{2}\;\right\rangle = \int_{0}^{s}  \int_{0}^{s}dt'\; dt'' \;\; \left\langle\; \theta(t')\;\theta(t'')\;\right\rangle .
\end{equation}
For large flight times (where $s$ in the upper bound of the integral is much greater than the correlation length scale), there is a simple connection between the variance of the fractional luminosity fluctuations and the variance of the expansion:  
\begin{equation}
\left\langle\;\left(\Delta L \over L\right)^{2}\;\right\rangle \propto \;  s^2 \left\langle\;  \theta^2(s)\;\right\rangle.\label{eq:dellum}
\end{equation}    

Let us show this for a typical positive-definite correlation function, $C_{1}(\Delta t)$.   We have   
\begin{eqnarray}
\left\langle\; \theta(t)\;\theta(t')\;\right\rangle& = &\; \;\int_{0}^{t} 
 dt_{1}\; \int_{0}^{t'}dt_{2} \;C_{1}(|\Delta t| ).
\end{eqnarray}
 The function $C_{1}$ is symmetric in its arguments, so just consider 
$t < t'$, then symmetrize later.  The first integration over $dt_{2}$ 
 produces the area under the Ricci correlation curve, $c_{o}$ . 
(We will be considering large flight times, so the edge effect at the origin 
is neglected.)  After the second integration, and symmetrizing,  we have the 
expression for the expansion correlation function:
 \begin{eqnarray}
 \left\langle\; \theta(t)\;\theta(t')\;\right\rangle& = & c_{0} \; t, 
\quad  t' > t,\label{eq:result1}\\
& =& c_{0} \; t', \quad  t > t'. \label{eq:result2}
\end{eqnarray}

Now we can calculate the fractional luminosity fluctuations:
\begin{eqnarray}
\left\langle\;\left(\Delta L \over L\right)^{2}\;\right\rangle& = &\int_{0}^{s}  \int_{0}^{s}dt\; dt' \;\; \left\langle\; \theta(t)\;\theta(t')\;\right\rangle \\
& = &2\; \int_{0}^{s} dt\; \int_{0}^{s} dt' \;\; \left\langle\; 
\theta(t)\;\theta(t')\;\right\rangle\;, \quad t > t', \\
& = &2\,  c_{0} \;\; \int_{0}^{s} dt\; \int_{0}^{t} dt'  \; t'\\
& = &\, c_{0}\; \int_{0}^{s} dt\;t^{2}\\
& = &{1 \over 3}  c_{0} \; s^{3}\\
& = &\,{1 \over 3}\;s^2 \left\langle \theta(s)^{2}\right\rangle.
\end{eqnarray}

Explicit calculations show that Eq.(\ref{eq:dellum}) also holds for the vacuum 
correlation function encountered in this paper.  In the case of the vacuum term, direct calculation reveals
\begin{eqnarray}
\left\langle\;\left(\Delta L \over L\right)^{2}\;\right\rangle_{vac}
& = &\,{1 \over 4}\;s^2 \left\langle \theta(s)^{2}\right\rangle_{vac}.
\end{eqnarray}

\section{}

In this Appendix, we will discuss two methods for defining quantities such as
$\langle 1/r^8\rangle$.
 The first approach involves an integration by parts. Let us first 
illustrate this approach in one space dimension with the quantity
\begin{equation}
\rho^{(1)}_n = \langle x^n \rangle = \int_{-\infty}^\infty dx\, x^n \,g_1(x)
\,,             \label{eq:one_dim_av}
\end{equation}
where $g_1(x) = {\rm e}^{-x^2/b^2}/(\sqrt{\pi}\, b)$. If $n > -1$, then
the above integral is absolutely convergent and there is no ambiguity in
the definition of $\rho^{(1)}_n$. If, however, $n < -1$, then the integral
in Eq.~(\ref{eq:one_dim_av}) diverges at $x=0$. In the case that $n$ is a
negative integer, let $m=-n$ and write
\begin{equation}
x^n = \frac{1}{x^m} = \frac{(-1)^{m-1}}{2\, (m-1)!}\, \frac{d^m}{dx^m}
\left[ \ln\left(\frac{x^2}{x_0^2}\right) + P_{m-1}(x) \right] \,,
\end{equation}
where $x_0$ is a constant and $P_{m-1}(x)$ is an arbitrary $(m-1)$-th degree polynomial 
in $x$. We can now insert this identity into Eq.~(\ref{eq:one_dim_av}) and
integrate by parts to define
\begin{equation}
\rho^{(1)}_n = \left \langle \frac{1}{x^m} \right\rangle = 
-\frac{1}{2\, (m-1)!}\, \int_{-\infty}^\infty dx\,
\left[ \ln\left(\frac{x^2}{x_0^2}\right) + P_{m-1}(x) \right]\;
\frac{d^m}{dx^m}\, g_1(x) \, .
\end{equation}
The resulting integrals are not only convergent, but are also independent of 
the choices of $x_0$ and the coefficients of $P_{m-1}(x)$. (These are not really
independent, as a shift in $x_0$ shifts the zeroth order term in $P_{m-1}$.)
Furthermore, the results are always of order $b^{-m}$. For example,
\begin{equation}
\langle 1/x^2 \rangle = -2/b^2\, ,
\end{equation}
\begin{equation}
\langle 1/x^4 \rangle = 4/(3b^4)\, ,
\end{equation}
\begin{equation}
\langle 1/x^6 \rangle = -8/(15b^6)\, ,
\end{equation}
and
\begin{equation}
\langle 1/x^8 \rangle = 16/(105b^8)\, .
\end{equation}

Note that if we were to allow the polynomial $P_{m-1}$ to be a function of $|x|$, 
rather than $x$, then the final result will depend upon the coefficients of
the odd powers of $|x|$. For example, let
\begin{equation}
P_1 = P_1(|x|) = c_0 +c_1\,|x| \,.
\end{equation}
Then we find
\begin{equation}
 \rho^{(1)}_{-2} =  -\frac{2}{b^2} - \frac{c_1}{\sqrt{\pi}\, b}\,.
\end{equation}
In effect, what has happened here is the addition of a term proportional
to $\delta(x)$ into the representation of $1/x^2$, which has caused
$\rho^{(1)}_{-2}$ to acquire a term proportional to $g_1(0)$ with an
arbitrary coefficient.

The second approach to defining the $\rho^{(1)}_n$ involves a recurrence 
relation. Note that if we let $\alpha = 1/b^2$, then we can write
\begin{equation}
\frac{\partial g_1}{\partial \alpha} = 
\left(\frac{1}{2 \alpha} - x^2 \right)\, g_1 \,.
\end{equation}
This allows us to derive the recurrence relation
\begin{equation}
\rho^{(1)}_{n+2} = - \sqrt{\alpha}\, \frac{d}{d \alpha}\,
 \left(\frac{\rho^{(1)}_{n}}{\sqrt{\alpha}} \right)\,. \label{eq:1D_recurr}
\end{equation}
We can start from $\rho^{(1)}_0 = 1$ and uniquely generate all $\rho^{(1)}_n$,
where $n$ is a positive even integer from Eq.~(\ref{eq:1D_recurr}). We can
also integrate this relation to define $\rho^{(1)}_n$ for negative even values of
$n$, but the result will contain undetermined constants of integration
which are equivalent to the arbitrary coefficients in $P_{m-1}(|x|)$.

Now we turn to two space dimensions and examine the quantities
\begin{equation}
\rho^{(2)}_n = \langle r^n \rangle = \int d^2x\, {r^n}\, g(r) =
2\, \pi\, \int_0^\infty dr\, r^{n+1}\, g(r)
\,,             \label{eq:two_dim_av}
\end{equation}
where the sampling function is now $g(r) = {\rm e}^{-r^2/b^2}/(\pi\, b^2)$.
Here the integrals converge if $n > -2$, but are divergent for
$n \le -2$. In the latter case, we can still use integration by parts to
define $\rho^{(2)}_n$. Consider first the case where $n=-2\ell$ is a negative, even
integer, and write
\begin{equation}
\frac{1}{r^{2\ell}} = (\nabla^2)^\ell 
\left[ a_\ell \, \ln^2\left(\frac{r^2}{r_0^2}\right) + f_\ell(r) \right] \, ,
                                            \label{eq:diff}
\end{equation}
where
\begin{equation}
\nabla^2 = \frac{1}{r}\, \frac{d}{dr}\,\left(r \frac{d}{dr}\right)
\end{equation}
is the Laplacian operator and $f_m(r)$ is any function which satisfies
\begin{equation}
(\nabla^2)^\ell \; f_\ell(r) = 0
\end{equation}
for $r \not= 0$. 
If we insert Eq.~(\ref{eq:diff}) into Eq.~(\ref{eq:two_dim_av}), the
Laplacian operators may be successively moved to act on $g(r)$ by repeated
use of Green's theorem:
\begin{equation}
\int (F \nabla^2 G -G \nabla^2 F) dV = \oint (F \nabla G -G \nabla F) da \,.
\end{equation}
Here the region of integration is a disk of radius $R$. In the limit that
$R \rightarrow \infty$, the surface term vanishes because $g(r)$ and its
derivatives vanish rapidly. The result is
\begin{equation} 
\rho^{(2)}_{-2\ell} = \left\langle\frac{1}{r^{2\ell}}\right\rangle =
2\, \pi\, \int_0^\infty dr\, r\,
\left[ a_\ell \, \ln^2\left(\frac{r^2}{r_0^2}\right) + f_\ell(r) \right] \,
(\nabla^2)^\ell \, g(r) \,.
\end{equation}
This integral is convergent, but depends upon various arbitrary constants.

This is best illustrated with some specific examples. First let $\ell =1$,
in which case we find $a_1=1/8$ and
\begin{equation}
f_1 = c_0\, \ln\left(\frac{r}{c_1}\right) \,.
\end{equation}
This leads to the result
\begin{equation}
 \left\langle\frac{1}{r^{2}}\right\rangle = 
- \frac{1}{b^2}\,\left[2 \ln\left(\frac{r_0}{b}\right) +\gamma + c_0\right]\,,
                                            \label{eq:r2}
\end{equation}
where $\gamma$ is Euler's constant. There is really one arbitrary constant
in this result, as $c_0$ may be absorbed into a redefinition of the constant
$r_0$. We can understand the appearance of $c_0$ in Eq.(\ref{eq:r2}) from 
the fact that 
\begin{equation}
\nabla^2\, f_1(r) = -2 \pi \,c_0\, \delta({\bf x})
\end{equation}
in two dimensions.
Similar results arise for larger values of $\ell$, but with more constants,
For $\ell=2$, where $a_2=1/32$, we can solve $\nabla^2 f_2 = f_1$ to find
\begin{equation}
f_2 = \frac{1}{4}\,c_0\,r^2\left[\ln\left(\frac{r}{c_1}\right) -1 \right]
+c_2\, \ln\left(\frac{r}{c_3}\right) 
\end{equation}
and 
\begin{equation}
\left\langle\frac{1}{r^{4}}\right\rangle =
\frac{1}{b^4}\,\left[2 \ln\left(\frac{r_0}{b}\right) +\gamma +1 -8c_2 \right]
+\frac{2 c_0}{b^2}\,.
\end{equation}
For $\ell=3$, $a_3=1/512$ and
\begin{equation}
f_3 = \frac{1}{128}\,c_0\,r^4\left[2\ln\left(\frac{r}{c_1}\right) -3 \right]
+ \frac{1}{4}\,c_2\,r^2\left[\ln\left(\frac{r}{c_3}\right) -1 \right]
+ c_4\, \ln\left(\frac{r}{c_5}\right)\,,
\end{equation}
which yields
\begin{equation}
\left\langle\frac{1}{r^{6}}\right\rangle =
-\frac{1}{4b^6}\,\left[4 \ln\left(\frac{r_0}{b}\right) +2\gamma +3 -64c_4 \right]
-\frac{8 c_2}{b^4}+\frac{2 c_0}{b^2}\,.
\end{equation}
Finally for $\ell=4$, where $a_4=1/18432$,
\begin{eqnarray}
f_4 &=& 
\frac{1}{13824}\,c_0\,r^6\left[6\ln\left(\frac{r}{c_1}\right) -11 \right]
+\frac{1}{128}\,c_2\,r^4\left[2\ln\left(\frac{r}{c_1}\right) -3 \right]
    \nonumber  \\
&+& \frac{1}{4}\,c_4\,r^2\left[\ln\left(\frac{r}{c_3}\right) -1 \right]
+c_6\, \ln\left(\frac{r}{c_7}\right) 
\end{eqnarray}
and
\begin{equation}
\left\langle\frac{1}{r^{8}}\right\rangle =
\frac{1}{36b^8}\,\left[12 \ln\left(\frac{r_0}{b}\right) +6\gamma +11 \right]
-\frac{768 c_6}{b^8} + \frac{64 c_4}{b^6} -\frac{8 c_2}{b^4}+ \frac{2 c_0}{b^2}\,.
\end{equation}

Now let us consider the case where $n=-2\ell-1$ is a negative odd integer
and write
\begin{equation}
\frac{1}{r^{2\ell+1}} = (\nabla^2)^\ell 
\left[ \frac{1}{[(2\ell-1)!!]^2\, r}  + f_\ell(r) \right] \, .
                                            \label{eq:diff2}
\end{equation}
A procedure analogous to that in the case of even $n$ leads to
\begin{equation} 
\rho^{(2)}_{-2\ell-1} = \left\langle\frac{1}{r^{2\ell+1}}\right\rangle =
2\, \pi\, \int_0^\infty dr\,
\left[ \frac{1}{[(2\ell-1)!!]^2} +  r\, f_\ell(r) \right] \,
(\nabla^2)^\ell \, g(r) \,.
\end{equation}
The results for $\ell=1$ and $\ell=2$ are
\begin{equation}
\left\langle\frac{1}{r^{3}}\right\rangle = -\frac{2\sqrt{\pi}}{b^3} 
+\frac{2 c_0}{b^2}
\end{equation}
and
\begin{equation}
\left\langle\frac{1}{r^{5}}\right\rangle =\frac{4\sqrt{\pi}}{3 b^5}
-\frac{8 c_2}{b^4} + \frac{2 c_0}{b^2} \,.
\end{equation}

We can also derive a recurrence relation for $\rho^{(2)}_n$. Note that
\begin{equation}
\frac{\partial g}{\partial \alpha} = 
\left(\frac{1}{ \alpha} - r^2 \right)\, g \,.
\end{equation}
This leads to the relation
\begin{equation}
\rho^{(2)}_{n+2} = - {\alpha}\, \frac{d}{d \alpha}\,
 \left(\frac{\rho^{(2)}_{n}}{{\alpha}} \right)\,. \label{eq:2D_recurr}
\end{equation}
We may start from either $\rho^{(2)}_{0}=1$ or $\rho^{(2)}_{-1}= \sqrt{\pi}/b$
and integrate this recurrence relation to find $\rho^{(2)}_n$ for all
negative integers $n$. The results agree with those of the integration
by parts approach, with the constants $c_i$ arising as constants of
integration.

Now we wish to address the difficult question of whether these constants can be
determined by physical arguments. Let us first consider the one-dimensional
case. So long as we require the polynomial $P_{m-1}$ to be a function of $x$,
as opposed to $|x|$, then
\begin{equation}
\frac{d^m}{dx^m}\, P_{m-1} = 0\, ,
\end{equation}
and there is no ambiguity. The undetermined constants which appear if 
$P_{m-1}$ is a polynomial in $|x|$, or if we integrate the recurrence relation,
all have dimensions of inverse powers of length.  One way to resolve the
ambiguity is to require that $\rho^{(1)}_n$ be proportional to $b^n$, which 
sets all of the constants of integration to zero. This is essentially a scaling 
requirement, that if $b \rightarrow 2\,b$, for example, then
$\rho^{(1)}_n \rightarrow 2^n\,\rho^{(1)}_n$. Another way to say this is that
$\rho^{(1)}_n$ should depend only upon the sampling scale $b$ and not on any
additional length scales.

For odd $n$, we can apply the same requirement to the two-dimensional case, and similarly
require that 
\begin{equation}
\rho^{(2)}_n \propto b^n \,.
\end{equation}
In the case that $n$ is a negative odd integer, this uniquely defines
$\rho^{(2)}_n$ by setting the constants $c_1 = 0$. For example, now
\begin{equation}
\left\langle\frac{1}{r^{3}}\right\rangle = -\frac{2\sqrt{\pi}}{b^3} 
\end{equation}
and
\begin{equation}
\left\langle\frac{1}{r^{5}}\right\rangle =\frac{4\sqrt{\pi}}{3 b^5} \,.
                                           \label{eq:r5}
\end{equation}
For $n$ even and negative, there is still an undetermined multiplicative 
constant associated with a term logarithmic in $b$, so we will have
\begin{equation}
\left\langle\frac{1}{r^{2 \ell}}\right\rangle = 
                              \frac{A_{2 \ell}}{b^{2 \ell}}\,,
\end{equation}
where we allow $A_{2 \ell}$ to possess a possible logarithmic dependence on $b$.

\end{document}